
%
\baselineskip=15pt
\magnification=1200
\hsize=15.2truecm
\vsize=23truecm
\overfullrule=0pt

\font\eightrm=cmr8
\parindent=8truemm
\parskip=4pt
\noindent
\centerline{\bf STAR FORMATION IN DWARF IRREGULAR GALAXIES:}
\centerline{\bf NGC~6822 $^*$}
\vskip 1.5truecm
\noindent
\centerline{G.Marconi$^1$, M.Tosi$^2$, L.Greggio$^3$, P.Focardi$^3$}
\vskip 1.5truecm
\par\noindent $^{1}$ European Southern Observatory, Garching bei Munchen,
Germany
\par\noindent $^{2}$ Osservatorio Astronomico, Bologna, Italy
\par\noindent $^{3}$ Dipartimento di Astronomia, Universit\`a di Bologna, Italy
\vskip 4.5truecm
\noindent
$^*$Based on observations made at the ESO telescopes in La Silla, Chile
\bigskip
\bigskip
\bigskip
\bigskip
\par\noindent Submitted to: The Astronomical Journal
\null\vskip 2.0truecm
$$\vbox{\settabs 3 \columns
\+ Send correspondence to: & M. Tosi &\cr
\+                         & Osservatorio Astronomico &\cr
\+                         & Via Zamboni 33 &\cr
\+                         & I - 40126 BOLOGNA &\cr
\+                         & ITALY &\cr
}$$
\vfill\eject\noindent
\centerline {ABSTRACT}
\medskip\noindent
Our proposed method to study the star formation histories in nearby
irregulars is here applied to NGC 6822. To this purpose we have obtained
accurate CCD photometry of
three regions in the galaxy, reaching V$\simeq$23.5 with the required accuracy
of $\sigma_{BVR}\leq$0.1 mag. Major information on the stellar populations,
star
formation rates and initial mass functions of these regions in the last 1 Gyr
are derived from the comparison of the observational color-magnitude diagrams
and luminosity functions with the corresponding synthetic diagrams and
luminosity functions generated by a numerical simulation code based on stellar
evolutionary tracks.
We find that in the last 1 Gyr the star formation activity has been rather
continuous, possibly occurring in long episodes of moderate activity separated
by short quiescent periods, and that the initial mass function exponent
$\alpha$
is in the vicinity of Salpeter's value, i.e. slightly flatter than in the solar
neighborhood. The studied regions of NGC~6822 appear to contain different
stellar populations. The preferred star formation regime for Region A is
a continuous activity coupled with a moderately flat $\alpha$. For Region C,
instead, our best models suggest a discontinuous
star formation and a moderately steep $\alpha$.
\par
NGC 6822 is the last object of our original sample of five nearby irregulars.
We can then draw the scenario resulting from this sample: most of the
observed fields appear to have experienced a {\it gasping} regime of star
formation, with average exponent $\alpha$=2.35 for the initial mass function.
Despite their small sizes, the majority of the sample galaxies contain
different stellar populations in different regions.
\bigskip\noindent
1.~INTRODUCTION
\medskip\noindent
In the last few years increasing attention is being devoted to dwarf galaxies
as they represent a unique tool to better understand the evolution of galaxies
of the various
morphological types. Dwarf irregulars, in particular, are being intensively
studied, on the one hand because they are poorly evolved systems and this
in principle allows an easier interpretation of their history, and on the other
hand because many of them are relatively close to us and thus easy to observe.
The star formation (SF) rates in irregulars have been estimated by several
authors and with various methods (e.g. Dennefeld $\&$ Tammann 1980, Hunter $\&$
Gallagher 1986, Kennicutt 1992 and references therein). The most direct
approach,
as mentioned by Kennicutt (1992), is to analyse directly the resolved stellar
population. In this framework, we have proposed (Tosi {\it et al.} 1991,
hereinafter Paper I)
a new method to derive the SF histories in nearby irregular
galaxies, based on deep and accurate CCD photometry coupled with
synthetic color-magnitude (C-M) diagrams. It represents an improvement
over the classical isochrone fitting because it allows to estimate not only the
stellar ages, but also the initial mass function (IMF) and star formation rate
(SFR) in better agreement with the observational data.
\par
The most suitable objects for this kind of project are the irregulars of the
Local Group, where deep photometric exposures can resolve also the fainter,
i.e.
older, stellar population. We can therefore have direct information on the
star formation activity over a long period of time.
The galaxies of our sample are listed in Table 1, together
with their coordinates, derived distance modulus, number of observed fields,
number of stars detected in all the observed bands, and number of selected
stars with photometric error $\sigma\leq$ 0.1mag. We have observed at least
a couple of fields in each galaxy in order to check whether or not these small
size systems can be considered as single homogeneous bodies. In each field
we have detected objects  down to a
$3\sigma$ level over the background noise. This criterium allowed to reach  a
calibrated limiting magnitude of $<$ 26.5 in the B-band.  For
sake of accuracy we have subsequently retained only the objects simultaneously
present
in B,V and R (when observed) bands,
 and  with photometric error smaller
than a tenth of a magnitude (see next section).
This further selection has restricted our limiting magnitude to $<$24.5.
 At the distances of our sample
galaxies, MS stars with this apparent magnitude have masses as small as $\sim
2-3$ M$_{\odot}$ and, correspondingly, MS lifetimes up to $\sim 0.5-1$ Gyr.
This value is therefore the time interval over which our investigation can
provide reliable estimates of the SF and IMF in the observed galactic regions.
\par
In Paper I we have described in detail our method and examined DDO~70
(Sextans B), whose two observed regions turned out to contain roughly the
same stellar populations. A similar analysis, published by Ferraro
{\it et al.} (1989), has revealed instead very different populations in the
two studied regions of DDO 221 (WLM). In
Paper II (Greggio {\it et al.} 1993) we have presented the results relative to
DDO~210$^1$
\footnote{}
{\noindent
$^1$ For an unfortunate combination of errors, in Paper II and in Marconi
et al. (1990) we have attributed to DDO~210 different coordinates from
those given by Fisher and Tully (1975) and claimed ours to be correct.
We apologize with J.R.Fisher and R.B.Tully and with the readers for our mistake
and thank H.G.Corwin for pointing it out.}
and DDO~236 (NGC~3109), and found also in the latter galaxy
different stellar populations associated to different fields. Here, we will
examine DDO~209 (NGC~6822) and draw some general conclusions on all the
galaxies of the sample.
\par
NGC~6822 $(\alpha_{1950}= 19^{h}42^{m}07^{s}$, $ \delta_{1950} =
-14^{\circ}55^{\prime}01^{\prime\prime}$, $l= 25.3^{\circ}$,
$b= -18.4^{\circ})$ is a well studied galaxy of type IB(s)m, the first
object to be actually recognized by Hubble (1925) as an external system.
Thanks to its relatively short distance (m-M)$_o\simeq$23.5, this galaxy has
been studied by several authors from the radio (e.g. Klein \& Grave 1986) to
the X-ray (e.g. Fabbiano, Kim \& Trinchieri 1991) bands. Its stellar content
is easily resolvable and has been analysed both photometrically and
spectroscopically. The first complete photographic study on NGC~6822 was
performed by Kayser (1967) who attributed to the galaxy 29 variables, of
which 13 classified as Cepheids. On the basis of the period-luminosity (P-L)
relation derived for these Cepheids, she inferred a distance modulus
(m-M)$_o$=23.75. McAlary {\it et al.} (1983, hereinafter McA)
observed her Cepheids in the infrared to reduce the uncertainties due
to the high and variable reddening E(B-V)$\geq$0.3 affecting this low latitude
galaxy, derived a new P-L relation and obtained (m-M)$_o$=23.47$\pm$0.11.
This modulus and a variable reddening E(B-V)=0.3-0.5 have been confirmed
and/or adopted in all subsequent studies. Hodge (1980) found 16 OB associations
and 31 star clusters. In addition, Hodge, Kennicutt \& Lee (1988) have
extensively studied the system in H$_{\alpha}$ and detected up to 157 HII
regions belonging to NGC 6822. The only two CCD photometries
published so far have been obtained by Hoessel \& Anderson (1986, hereinafter
HA), who covered all the galaxy down to V$\simeq$22, and by Wilson (1992),
who searched for blue and red stars down to V $\simeq$ 23. Also Gallart {\it
et al.} (1994) have very recently studied the stellar content of NGC 6822
on the basis of very deep CCD frames in the V, R and I bands. They succeed in
detecting a well populated Asymptotic Giant Branch and interpret this feature
as a signature of a burst of SF occurred between 1.4 and 1.9 Gyr ago.
\par
NGC~6822 is an interesting object also from the spectroscopic point of
view. From seven of its HII regions
Pagel, Edmunds \& Smith (1980) derived an average oxygen abundance of
12+log(O/H)=8.25, a very low log(N/O)=-1.69
ratio and no evidence for abundance
gradients. Aaronson, Mould \& Cook (1985) and Aaronson, Cook \& Norris
(1985) analysed its C and M star content, found one S star and consequently
assigned to NGC~6822 a chemical abundance slightly larger than that of the
SMC, in agreement with Pagel's {\it et al.} A similar conclusion was reached
by Elias \& Frogel (1985) who studied more than 15 red supergiants and by
Azzopardi, Lequeux \& Maeder (1988) who derived a metallicity Z=0.0045 from an
analysis of the Wolf-Rayet stars.
\par
In the following we describe the data acquisition and reduction
 and present the resulting C-M diagrams
(Sect.3) and luminosity functions (Sect.4). These data are compared in Sect.5
with the corresponding predictions of the simulation procedures described
in Paper I. The results on NGC 6822 are examined in Sect.6 while a general
discussion on all the galaxies of our sample is presented in Section 7.
\bigskip\noindent
2.~THE DATA
\bigskip\noindent
2.1~{\it Observations}
\medskip\noindent
All the observations have been carried out with an RCA-CCD mounted
at the 2.2~m
ESO-MPI telescope in La Silla, with a frame size of
1.8$^{\prime}\times$2.9$^\prime$ and a pixel size corresponding to
0.17$^{\prime\prime}$/pxl. All the fields have been observed in the B and V
Johnson's standard filters and only two fields also in the R Cousins' band.
The individual exposures were up to a maximum of 30$^m$ in the V and R bands
and 70$^m$ in B.
\par
Three partially overlapping fields on the main body of NGC~6822 were observed
during three photometric nights in July 1991. The fields were chosen on the
basis of Hodge's {\it et al.} (1988) survey in order to avoid the presence of
HII regions which would have introduced much higher difficulty and uncertainty
in the data reduction. They are centered at
\par\noindent
$\alpha_{1950} = 19^{h} 42^{m} 07^{s}$, $\delta_{1950} = -14^{\circ}
58^{\prime}
 24^{\prime\prime}$, \par\noindent
$\alpha_{1950} = 19^{h} 42^{m} 15^{s}$, $\delta_{1950} = -14^{\circ}
55^{\prime}
 54^{\prime\prime}$, \par\noindent
$\alpha_{1950} = 19^{h} 42^{m} 07^{s}$, $\delta_{1950} = -14^{\circ}
53^{\prime}
 23^{\prime\prime}$; \par\noindent
and in the following they will be referred to as Region A, Region B and Region
C, respectively. The first and third fields are located along the optical bar
of the galaxy, whilst Region B is centered approximately 2$^{\prime}$ off
the bar. During the same nights an external field
located at $\alpha_{1950} = 19^{h} 41^{m} 24^{s}$, $ \delta_{1950} =
 -14^{\circ} 56^{\prime} 01^{\prime\prime}$  has been observed in order to
estimate the field contamination in the direction of the galaxy. Figure 1
shows the position of our fields on an enlargement of the blue print of the
Palomar Sky Survey. Eight primary photoelectric calibrators (Landolt 1983) and
ten secondary ones (Alcaino, Liller \& Alvarado 1987) spanning over a wide
color
range have also been observed to guarantee a good photometric calibration.
The journal of the observations indicating the number of frames available in
each band for each region, the exposure time and seeing conditions is reported
in Table 2. Note the very good seeing (relative to our site and telescope) of
$0.81^{\prime\prime}$ reached on July 12 1991.

\bigskip\noindent
2.2~{\it Reduction Procedures}
\medskip\noindent
The data have been reduced with the DAOPHOT package (Stetson 1987), which makes
use of an automatic algorithm for star search above a
detection threshold limit. The sky is measured locally and an
instrumental magnitude is derived by means of a semiempirical PSF.
\par
The conversion of our instrumental magnitudes to the Johnson standard system
has been performed by means of the observed primary standard stars
(Landolt 1983), to which we added secondary calibrators (Alcaino {\it et al.}
1987) located in globular clusters (NGC~4590, NGC~4833 and NGC~5946).
The wide color
range spanned by these calibrators allows us to derive a good relation between
instrumental and standard magnitudes. Standard average absorption coefficients
for La Silla (Rufener 1986)
have been used to correct our data for the atmospheric extinction.
A plot of the residuals versus color for our standard stars together with
the best linear fit to the data
is shown in Figure 2.
\par
The transformation between the magnitude obtained with Daophot PSF fitting
and that with aperture photometry has been computed selecting a sample
of bright {\it isolated} stars in each frame.
The number of selected stars for each region ranged typically between 5 and 10.
An aperture magnitude was computed for all these stars and the zero point
term relating this aperture magnitude with the PSF one derived.
The dispersion
of this term amounted typically to a few hundredths of magnitude
and turned out to be the major contributor to the total photometric error,
since it is one order of magnitude larger than both the internal error
of the aperture photometry and the frame to frame zero point difference.
\par
All the selected stars have been measured with the same aperture.
Testing different aperture sizes we have estimated the internal error in the
aperture photometry to be, on average, less than 0.005 mag.
The small overlap between the  observed fields allowed us to estimate
the transformation between the instrumental magnitudes of different frames,
so as to avoid the introduction of spurious differences in the relative zero
points in the magnitude scales. As a further check,
we have independently calibrated each single frame and measured
the magnitudes of the stars belonging to the overlapping regions.
 For these stars we found $\Delta mag \leq 0.002$.
 The final derived equations, which relate our instrumental magnitudes
to the Johnson photometric system for Region A are:\par
\bigskip
         B = b + 0.112 ($\pm 0.011) \times $(b-v) $- 4.508$ ($\pm 0.05$)
\bigskip
         V = v + 0.016 ($\pm 0.011) \times $(b-v) $- 4.015$ ($\pm 0.05$)
\bigskip
         R = r + 0.015 ($\pm 0.020) \times $(v-r) $- 4.601$ ($\pm 0.04$)
\bigskip\noindent
where B, V and R are the magnitudes in the standard system,
and b, v and r are the instrumental magnitudes. Similar coefficients are
derived for the other regions.
\par
Due to the shorter exposure times, the CCD frames on NGC 6822 are not as deep
as those obtained for the other galaxies of our sample. The 2031 objects
detected in the three fields of NGC~6822 are in fact all brighter than V
$\simeq$ 24.5.
Of these, 480 were found in Region A and 452 in Region B in all the B, V and
R bands and 1099 in Region C only in the B and V bands for lack of
sufficient observing time. The standard
errors $\sigma$ derived with DAOPHOT for Region A  are displayed in Figure 3
as functions of the calibrated magnitudes. We recall that the
standard error provided by DAOPHOT
is not the "true" total photometric error. We have used the
short exposure frames  to measure the frame to frame error , that is
the rms scatter from the mean
value in each band (see Romeo et al. 1989). We found that, at least for
the bright stars,
it does not exceed 1.2 $\sigma$ (see also Stetson \& Harris 1988).
\par
To guarantee the good photometric accuracy required for a proper
interpretation
of the C-M diagrams and for a reliable derivation of the luminosity functions,
we have further selected the resolved stars and retained only those with
standard error $\sigma$ smaller than, or equal to, 0.1 mag.
This selection leaves us with a total of 1772 stars with V$\leq$23.5
(436, 395 and 941 in Regions A, B and C, respectively).
It is interesting to note that, thanks to the better seeing conditions
and to the smaller pixel size of the CCD used for these observations, the
errors derived for NGC 6822 are considerably smaller than those for Sextans B,
NGC 3109 and DDO 210 (see Papers I and II). This explains why the above
selection is less severe in this case and allows to keep 87$\%$ of the detected
stars, compared to 53$\%$ in Sextans B, 41$\%$ in NGC 3109 and 51$\%$ in DDO
210
(cfr. Table 1).
\par
Table 3 lists the positions, V magnitudes, B-V and V-R (when observed) colors
for all the objects selected in the three regions of NGC 6822 in all the
observed bands and with $\sigma$ smaller than 0.1 mag. The positions
are expressed in X and Y pixel coordinates, referred to the bottom left corner
of each frame.
\par
The completeness of our data as a function of magnitude has been determined
with particular care using the ADDSTAR routine in DAOPHOT. For each frame,
we have added in each half magnitude bin a different number of artificial
stars (10$\%$ of the stars detected in the same bin in the
original frame)
so as to avoid overcrowding (see Mateo 1988). The frames have
then been re-reduced and the completeness factors estimated
by checking how many
of the added stars were lost, either because undetected, or because affected
by a large photometric error($\sigma \geq 0.1 mag$).
The completeness was estimated averaging the results obtained running 10 times
ADDSTAR on each frame changing, on each run, the spatial distribution of the
stars.
The results for our three regions are reported
in Figure 4 and we wish to emphasize that the higher photometric
accuracy allows these data to remain complete for a larger magnitude interval
than in the other galaxies of our sample, but
leads also to a much more drastic
breakdown. The latter is also due to the large reddening that cuts down the
faint blue objects. Since we retain only stars detected with small error in
{\it all} the observed bands, the non detection in the blue band implies,
in fact, the complete loss of a substantial fraction of the objects.
\par
We have also analysed in detail what fraction
of the artificial stars is lost, or is recovered with strongly altered values,
because of a total or partial blending with other objects or with background
noise peaks. These cases of blending have significant effects on the derived
C-M diagrams and in particular on their color distribution: usually the
combination of blue and red stars leads to a {\it yellowing} of the resulting
color, but there are cases of blend occurring only in one band that yield a
spurious color extremely blue or extremely red. These cases are also partially
responsible of the blue and red faint tails in the observational C-M diagrams
of
resolved crowded galaxies.
\bigskip\noindent
2.3~{\it Comparison with Previous Photometry}
\medskip\noindent
   As already outlined in the Introduction, stellar photometry for NGC 6822
   has been presented by various authors.
   Kayser (1967) has published an UBV photographic photometry;
   subsequentely HA have obtained CCD g,r,i (Thuan \&
   Gunn 1976) photometry for the main body of the galaxy and, more recentely,
   Wilson (1992) has published CCD BV photometry for its blue and red stars.
\par
   HA compared their results with Kayser's, finding
   a good agreement in the V (even though their photometry was 0.1 mag
   fainter for stars with V$>$18) and a non linear trend in the B magnitudes.
   We have applied the transformations given by HA from the
   Gunn to the Johnson system and compared our magnitudes to theirs on a
   sample of about 70 common stars.  These common stars have been selected
   in regions of moderate crowding conditions.  We have decided to avoid
   extremely crowded fields where incorrect sky subtraction and
   blending of multiple objects are likely to occur and alter the comparison,
   especially because of the rather different seeing conditions (HA average
   seeing having been 1.5$^{\prime\prime}-2.0^{\prime\prime}$).
The comparison between our and HA's B, V and R magnitudes for the common stars
is shown in the top panels of Fig.5. Despite the scatter, the agreement is
fairly good, with perhaps a very weak trend of our photometry to be brighter
than
theirs at the bright end and fainter at the faint end. This small effect can
have a natural explanation in the different seeing conditions of the two
sets of observations. Larger seeing in crowded fields may cause, in fact,
an underestimate of the magnitude of bright stars due to an overcorrection
of the sky and an opposite effect on the fainter objects due to stellar blend.
\par
More recent B and V CCD photometry for the blue and red stars belonging to
NGC 6822 has been presented by Wilson (1992).
A computer readable version of Wilson's data, kindly made available  to us by
herself, allowed the running of a search algorithm on both (ours and hers)
data and we could thus find about 100 common objects.
The comparison of the two photometries is shown in the bottom panels
of Fig.5.
As it can be seen the consistency between the two photometries is rather
good but there is evidence for a zero point shift in both bands. We have no
obvious explanation for such a shift which amounts roughly
to 0.15 in B and 0.1 in V, our photometry being fainter on both bands and
on the whole magnitude range. A possible explanation for this discrepance might
be found in the numerous observational problems mentioned by Wilson and
by the related calibrations and recalibrations she had to perform.
Since the shifts in the two bands are similar
to each other, her resulting colors are in agreement with ours and so is
the color distribution of the stars in the C-M diagram.

\bigskip\noindent
3.~COLOR-MAGNITUDE DIAGRAMS
\medskip\noindent
For internal consistency and homogeneity, hereinafter we will present for all
the observed fields all the objects selected in the B and V bands, although for
Regions A and B the R magnitudes are also available. For these two regions, we
have actually applied all the tests also to the stars further selected in R and
found no different result from those described below. What obviously changes
is the number of studied stars , which is larger because selecting only in B
and
V is less restrictive. We think, however, that a homogeneous selection allows
for a more direct comparison between the derived properties of the different
regions.
\par
Fig.6 shows the C-M diagrams of the studied regions of NGC 6822 resulting from
the data described in the previous section. All the three diagrams of Fig.6
refer to stars selected in B and V and contain 602, 452 and 941 objects for
Regions A, B and C respectively. The numbers corresponding to Regions A and B
are larger that those quoted in Sect.2.2 because of the less restrictive
selection criterion. All the plotted objects have $\sigma\leq$0.1.
The lower total number of
objects in Region B may be due to its location in the galaxy. Region
B is in fact the only field located well outside the central bar where most
of the bright stars of NGC 6822 seem to be concentrated. Both Regions A and C
are located on the bar and the latter field is quite close to the galaxy
center. This central position may be the origin of the larger stellar content
(and activity, as we shall see in Sections 5 and 6) of Region C.
\par
Fig.7 shows in turn the C-M diagram of the external field. The importance of
this
additional field is apparent: by comparing Figs 6 and 7 it is clear that the
bright vertical sequence extending in all the diagrams around B$-$V$\simeq$0.9
is most probably populated by foreground stars. Conversely, the remaining
portions of the diagrams look virtually unaffected by contamination.
\par
The general morphology of the C-M diagrams of NGC 6822 is similar to that of
any dwarf irregular (see e.g. Aparicio {\it et al.}
1987, Freedman 1988, Papers I and II, Bresolin, Capaccioli \& Piotto 1993),
with a quite scattered
distribution and a prominent concentration of stars in the so-called blue
plume. The colors, however, are significantly redder than in the other
galaxies of our sample; for instance, the median color of the blue plumes
of Fig.6 is B-V$\simeq$0.1 whereas all the corresponding colors of the other
galaxies are negative and range between -0.1 and -0.2. This difference
is essentially due to the much larger reddening affecting NGC 6822. In fact,
the most popular value for it is E(B-V)=0.36 (McA),
while the reddenings for Sextans B, NGC 3109 and DDO 210 are all around 0.03.
In addition, several authors (e.g. van den Bergh and Humphreys 1979,
Humphreys 1980, McA) have found that in NGC 6822
the reddening is highly variable and may reach values as high as E(B-V)=0.6.
This internal variation may be the cause of the color shift of the C-M
diagram of our three observed regions as apparent in Fig.6, the blue edge
of the blue plume becomes progressively redder from Region C to
Region A to Region B. This color shift cannot be attributed to frame to frame
zero point magnitude variations, as discussed in section 2.2 .
In the following we will adopt McA's
E(B-V)=0.36 for Region C and, consequently, E(B-V)=0.4 for Region A and
E(B-V)=0.45 for Region B.
\par
The morphology of the blue plume of Region C requires one further comment.
As already noticed for other irregulars (e.g. Freedman 1988, Papers I and II),
their blue plumes are
populated not only by main sequence (MS) stars but also by stars at the
blue edge of the blue loop evolutionary phase, and the two populations
appear mixed because of the image blend occurring on the CCD frames.
This represents a serious warning for authors deriving
the IMF from the luminosity functions of the blue plume as if it contained
only MS stars. In absence of any stellar blend, the morphology of the
evolutionary
tracks is such that
a vertical gap should appear in the blue plume separating
MS and post-MS stars (see e.g. fig.2 and relative discussion in Tosi 1994).
Since the frames relative to Region C have been taken in better seeing
conditions than the others (see Table 2), the photometry is less affected by
stellar blending. Indeed, the blue plume stars brighter than V$\simeq$19.5 in
the bottom panel of Fig.6 seem somehow splitted in two vertical sequences,
bluer and redder than B-V$\simeq$0.15: is this the signature of the two
distinct
evolutionary phases~? And would the separation be actually significant in
even better seeing conditions~? To answer these questions we should wait
for observing time on telescopes with higher performances and much better
average seeing.
\bigskip\noindent
\bigskip\noindent
4. LUMINOSITY FUNCTIONS
\medskip\noindent
The differential luminosity functions relative to the stars selected in the
B and V bands in the three regions of NGC 6822 are shown in the top panel of
Fig.8. The brightest portions of the curves tend to overlap because they are
dominated by the population of foreground stars; the differences between the
remaining portions reflect instead the intrinsic properties of the three
fields.
The dashed curve corresponds to Region B and lies below the others because of
the significantly lower number of detected stars. Region A contains as well a
lower total number of stars than Region C, but its LF (dotted line) shows that
this underpopulation is limited to objects with 20.5$\leq$V$\leq$23. This, in
turn, means that the fraction of bright stars is proportionaly larger in Region
A than in Region C.
\par
The LFs relative to the MS stars of the three regions are shown in the bottom
panel of Fig.8. As recalled in the previous section, it is important to
distinguish MS stars from the evolved objects populating the blue plumes.
To this end, we have applied the criterion of Papers I and II, based on the
synthetic C-M diagrams which will be described in the next section.
In practice, we have considered as MS stars in Region A only the 183 objects
with -0.26$\leq$ B-V $\leq$0.24, in Region B the 131 objects with
-0.21$\leq$ B-V $\leq$0.29, and in Region C the 307 objects with
-0.30$\leq$ B-V $\leq$0.20. The different B-V ranges adopted for the MS
selection in the three fields are due to the different reddenings affecting
each region and mentioned in the previous section and the corresponding
different colors of the blue plume.
\par
The number of MS stars in Region C is significantly higher than that in the
other regions: this is mostly due to the larger total number of stars detected
in this field. In fact, the fraction of MS/total objects does not vary much,
and is 31$\%$ in Region A, 29$\%$ in Region B, and 33$\%$ in Region C. We
notice, as for the global LFs, that the bright portion of the dotted line of
Region A overlaps or even lies above the solid line of Region C, whereas at
magnitudes fainter than V=20 the dotted curve remains always below the solid
one. This indicates again that the fraction of bright MS stars is
proportionally
larger in Region A than in Region C.
\par
The slopes of the LFs are usually derived with least squares fitting to
their complete portions. However, we have shown in Paper II
that the large statitical fluctuations due to the small number of sampled
stars, specially in the brightest magnitude bins, may significantly alter
the results of a method requiring the data binning. To overcome this problem,
we have therefore preferred to apply a maximum likelihood fitting to evaluate
the slopes of the MS LF in NGC 6822. Down to V=21, where all the three regions
are fully complete in all the observed bands, the resulting slopes are:
$\Delta log N/ \Delta$ V = 0.47 $\pm 0.07$ for the 47 MS stars of Region A
brighter than the adopted magnitude limit,
$\Delta log N/ \Delta$ V = 0.58 $\pm 0.12$ for the 21 MS stars of Region B and
$\Delta log N / \Delta$ V = 0.62 $\pm 0.05$ for the 69 MS stars of Region C.
Therefore, Region C contains a larger fraction of MS stars but relatively less
bright objects than the other fields. We will discuss this point in Section 6
and see that it can be explained on the basis of the IMFs and SFRs derived
from our simulations. If we sum up the data for all the three regions to
have a larger sample and an average MS LF for NGC 6822, the resulting slope
is 0.55 $\pm 0.05$.
\par
Our slopes on average are consistent with those of the other galaxies of our
sample (Papers I and II) and with those derived by Freedman (1985) and by
Hoessel (1986). However, the slope of Region A appears flatter than the others
and comparable only to that of Region 1 in WLM (Ferraro {\it et al.} 1989).
We will see in the next Section that this may indeed correspond to a flatter
slope of the region IMF.
\bigskip\noindent
5.~THEORETICAL INTERPRETATION
\medskip\noindent
The data described so far can be interpreted by means of theoretical
simulations
based on the stellar evolution theory. The procedure for generating synthetic
C-M diagrams and the corresponding LFs has been described in detail in Paper I.
Here we simply recall that it provides C-M diagrams with the same number of
stars as observed in each galactic field and takes into account all the
parameters (age, metallicity, IMF and SFR) affecting the stellar distribution
in
the C-M diagram. In addition, our simulations account for the stochastic nature
of the star formation processes, the small number statistics, the spread due
to photometric errors and image blend, and the incompleteness of the data
due to both the limited sensitivity of the instrumental setup and stellar
blend. The effects of the stellar blend in the CCD frames on the synthetic
diagrams and derived SFR and IMF have been widely examined in Papers I and II;
in the following we therefore present only the results of simulations taking
the empirical blend into account.
\par
The simulation code is based on complete sets of homogeneous stellar evolution
tracks of various metallicities. Homogeneous sets of models must be adopted
to avoid the appearance in the C-M diagrams of spurious features due to the
interpolation between inconsistent tracks. Our simulations have been
performed adopting the evolutionary tracks
computed by the Padova group (namely those
with metallicity Z=0.001 and large overshooting presented by Greggio 1984 and
Bertelli {\it et al.} 1986, hereinafter BBC86; the tracks with Z=0.008 and
moderate overshooting published by Alongi {\it et al.} 1993, hereinafter BBC93;
and the tracks with Z=0.008 and no overshooting presented in the same paper,
hereinafter BBC93CL).
\par
A large number of simulations have been performed with each of these sets of
stellar evolution models, varying IMF, SFR, starting epoch and duration of the
star formation activity. The IMF is assumed to follow a power law
m$^{-\alpha}$.
The SFR can be exponentially decreasing with time or constant, either in the
whole range of 1 Gyr or in smaller time intervals. In all cases we have adopted
McA's distance modulus (m-M)$_o$=23.47 and the reddening attributed to each
field in Section 3. In the following we examine only the two most populated
Regions (A and C) and present only the most relevant models.
\medskip\noindent
{\it 5.1~Region A}
\medskip\noindent
Figure 9 shows the synthetic diagrams based on the three different sets of
stellar evolution tracks which are in better agreement with the observational
data. The vertical sequence present in the observational C-M diagram and
corresponding to the foreground contaminating stars is obviously missing in
all the synthetic diagrams. The reddening applied
to the theoretical magnitudes to convert them to the empirical values is
E(B-V)=0.40 as derived in Section 3.
The diagram in panel (c) corresponds to the BBC86 models whose metallicity
Z=0.001 is slightly lower than that (Z=0.004) attributed to
NGC 6822 on spectroscopic bases (see the Introduction). In spite of this,
the synthetic blue plumes have colors in agreement with the observed ones.
The diagram in panel
(b) corresponds to the BBC93 tracks and that in panel (a)
to the BBC93CL tracks. Both sets assume a metallicity slightly larger than
that estimated for our system and attribute too red colors
to the synthetic stars. In addition to the reddening, we have therefore
applied to the synthetic colors a shift $\Delta$(B-V)=-0.03,
so as to obtain for the blue plume of the theoretical
distribution the same color of the observational one. The applied
color shift is consistent with the corresponding
expectations for MS stellar evolutionary models with Z varying from 0.008 to
0.004. In spite of this correction, the colors of the synthetic red supergiants
remain slightly too red.
\par
As far as the evolutionary phase of the stars in the blue plume is concerned,
we notice that different tracks provide different situations. In panel (c),
despite the inclusion of stellar blend in our simulations, MS and post-MS
stars are well separated by the Hertzsprung gap (at B-V$\simeq$0.2 in the
brighter portion of the plume). We should emphasize, however, that to avoid
the confusion due to too many parameters, our simulations have not taken into
account the probable presence of a significant fraction of physical binary
stars which could well fill the gap as the C-M diagrams of galactic open
clusters demonstrate (see e.g. Bonifazi {\it et al.} 1990 on NGC 2243).
In panel (b) the shape of the blue loops is
such that the gap is more evident at fainter magnitudes but undistinguishable
at the bright end where MS and post-MS stars are completely mixed. In panel (a)
the {\it nose} of the blue edge of the loops of massive stars is so evident,
with its vertical ridge line at B-V=0.3 and horizontal sequence at V=20, that
it is easy to distinguish MS and post-MS stars. This particular feature may
however be characteristic only of the adopted value of Z and is certainly very
sensitive to the input physics of the stellar models: a slightly bluer
extension
of the loops would be sufficient to let MS and post-MS stars merge.
\par
The diagram of panel (a) assumes an IMF exponent $\alpha$=2.2 (slightly
flatter than Salpeter's 2.35) and two episodes of SF activity, the older
one starting 1 Gyr ago and lasting 9$\times 10^8yr$ with a SFR =
5.6$\times 10^{-3}M_{\odot}yr^{-1}$, and the younger from 8$\times10^7yr$ ago,
lasting 7.6$\times10^7yr$, with a SFR = 2.8$\times 10^{-3}M_{\odot}yr^{-1}$.
The
last activity must have stopped 4 Myr ago to avoid the presence in the
synthetic
C-M diagram of bright stars not observed in the empirical diagram.
The model of panel (b) assumes Salpeter's IMF and two episodes of SF activity,
the older one from 1 Gyr ago, with a SFR = 2.8$\times 10^{-3}M_{\odot}yr^{-1}$
for 9$\times 10^8yr$, and the younger from 9$\times10^7yr$ ago
and lasting 8.8$\times10^7yr$ with the same SFR. The same result can
obviously be obtained with one single episode lasting from 1 Gyr to
2$\times10^6yr$ ago and the same SFR. Finally, the diagram of panel (c)
assumes a flatter IMF ($\alpha$=2.0) and a constant
SF activity from 1 Gyr to 2$\times 10^6yr$ ago but with a lower SFR =
1.3$\times 10^{-3}M_{\odot}yr^{-1}$.
\par
The goodness of our models with respect to the observational data can also
be evaluated by comparing the predicted and observed LFs.
The global and MS luminosity functions derived from the models of Fig.9 are
compared in Fig.10 with the corresponding observational distributions.
Notice that for a more direct comparison with the data, the synthetic
global LFs shown in the figure include the effect of the
foreground stars, as measured in the external field.
The simulations indicate that the B-V range populated only by MS stars is
-0.26$\leq$ B-V $\leq$0.24. We have then selected the objects for the MS LF
on the basis of this color criterium. All
the models reproduce very well the global LF of Region A. Also the MS LFs
are in agreement with the data, specially those corresponding to the BBC93
and BBC93CL tracks. The slight excess of bright stars in the dotted line
distribution is probably due to the much larger overshooting of the BBC86
tracks ($\lambda$=1,compared with $\lambda$=0.25 of the BBC93 and $\lambda$=0
of the BBC93CL) which implies larger luminosities for stars of the
same initial mass.
\par
As shown in Figs 9 and 10, the C-M diagrams and  the LFs resulting from
different stellar evolution models slightly differ from each other, as well as
the individual epochs of the SF episodes, the corresponding rates and the
derived IMF slopes. This is due to the different characteristics of the
adopted tracks. It is remarkable, however, that despite the uncertainties
related to the stellar evolution models, all the three best simulations
suggest for Region A a moderately flat IMF, a SFR ranging between 1 and
6 $\times 10^{-3}M_{\odot}yr^{-1}$ (possibly interrupted for a very
short time around 100 Myr ago) and no activity in the last few Myr.
This indicates that, beyond the precision of the derived values, the method
provides reliable information on the evolutionary scenario of the studied
regions.
\par
We have extensively tested the effects of varying the SF and the IMF.
{}From all the simulations performed on Region A we derive that a SFR
constant throughout the entire last 1 Gyr is inconsistent with
the observed stellar population because it provides too many MS stars whatever
the IMF. If the IMF is flat these overabundant MS stars populate mostly the
brighter portion of the sequence, otherwise they are concentrated on its
fainter
portion; in any case the shape of the MS LF is quite different from the
observational one. As discussed above, with the BBC86 and BBC93 tracks this
inconsistency is overcome if the SF activity stops a few Myr ago, but with the
BBC93CL models no satisfactory agreement has been reached with a constant SF.
\par
The latter tracks provide instead a good result if the SF has been
exponentially decreasing with time. Fig.11 shows the C-M diagram of a model
assuming a decreasing SF with e-folding time of 0.6 Gyr and average rate 8.1
$\times 10^{-3}M_{\odot}yr^{-1}$. The corresponding LFs are represented by
the thin solid curves in Fig.10. Notice that this kind of SF is still active
at the present time. In this case the IMF should be Salpeter's ($\alpha$=2.35),
because with a flatter slope the MS turns out overpopulated, specially at its
bright end, and with a steeper slope there are not enough luminous stars.
With the BBC93 models, an exponentially decreasing SF may also
be consistent with the data if combined with Salpeter's IMF (or with a steeper
slope $\alpha$=2.6) but the comparison of both the C-M diagrams and the LFs is
not completely satisfactory. With the BBC86 tracks, instead, there is no way
of reproducing the observed stellar distribution with this kind of SF, because
if the IMF is flat the MS turns out too bright and with too straight a vertical
shape, and if the IMF is steeper the MS is underpopulated.
These different results can be explained as follows:
The main consequence of a time decreasing star formation is
that it implies relatively more evolved stars than a constant SFR. This
characteristic may favor or not the agreement with the data, depending on the
morphology and luminosity of the adopted tracks, and on the relative lifetimes
of the various evolutionary phases. The BBC93CL tracks
cannot provide a satisfactory reproduction of
the observed stellar distribution if a constant SF is assumed whereas an
exponentially decreasing SFR compensates their larger fraction of MS stars and
allows to reach a good agreement. Vice versa, the BBC86 tracks predict a low
ratio of MS/post-MS stars and cannot be combined with a SF regime enhancing
this deficiency.
This trend may seem counterintuitive, since for a given mass models including
convective overshooting are characterized by longer core hydrogen burning
lifetimes and larger ratios between hydrogen and helium burning lifetimes.
The first of these properties is infact responsible for the lower SFR derived
for the best models going from BBC93CL to BBC93 to BBC86 data sets. However,
due to the strong evolution of the bolometric correction along the evolutionary
track of massive stars, at a given V magnitude the ratio between MS
and evolved objects does not reflect directly the ratio between MS and post-MS
lifetime for the same stellar mass. The combination of all the parameters
controlling the stellar distribution in this bright portion of the CMD results
into a lower global MS/post-MS objects for the tracks including convective
overshooting, once the level of the SFR has been fixed to reproduce the
correct total number of objects populating the CMD.
\par
As shown in Fig.9 a discontinuous SF provides excellent results both with the
BBC93 and the BBC93CL models. With the first set of tracks
the IMF slope should not be
steeper than Salpeter's to avoid an underpopulated upper MS and should not
be flatter for the opposite reason. The SFR is the same in the two episodes and
this (combined with the good result of the constant SFR case) suggests that
with these tracks the best SF regime for Region A is a constant rate stopped
some Myr ago. With the BBC93CL set and two episodes of SF, the best IMF is
slightly flatter ($\alpha$=2.2) and the SFR decreases significantly from the
first to the second episode. Combined with the good performance of the
exponentially decreasing models, this suggests that with this set of tracks
the best SF regime for Region A is a time decreasing activity.
Finally, with the BBC86 tracks a discontinuous SF does not provide very good
results, the only acceptable case being with Salpeter's IMF and a SFR larger
in the recent than in the older episode.
\bigskip\noindent
{\it 5.2~Region C}
\medskip\noindent
Figure 12 shows for each adopted set of stellar evolution models one of the
synthetic diagrams in better agreement with the data. The reddening adopted in
this region is McA's E(B-V)=0.36. In addition (see Section 5.1), we have
applied
a color shift $\Delta$(B-V)=-0.03 to the BBC92 and BBC92CL models. These values
let the observed and theoretical blue plumes have the same location in the
C-M diagram, while the distribution of the red supergiants remains slightly
too red, as already noticed for Region A.
\par
The model of Fig.12(a) is based on the BBC93CL tracks and reproduces fairly
well the empirical stellar
distribution. The only features of this synthetic diagram significantly
different from its observational counterpart are the Hertzsprung gap and the
lack of the vertical sequence at B-V$\simeq$0.8. We recall that the latter
corresponds to
foreground objects (see the external field in Fig.7) and must not be expected
in the synthetic diagram. The former is indeed an inconsistency of the models,
probably resolvable by taking into account the presence of binary stars.
The model assumes a moderately steep ($\alpha$=2.6) IMF and two episodes of
SF: one started 1 Gyr ago and lasted 9$\times10^8yr$ at a rate of
$2.2 \times 10^{-2}M_{\odot}yr^{-1}$ and the second started 9$\times10^7yr$
ago and stopped 3$\times10^6yr$ ago at a lower rate of
$8.0 \times 10^{-3}M_{\odot}yr^{-1}$. The brightest blue stars are both MS
and blue loop stars born in the recent SF episode.
\par
The model of Fig.12(b) is based on the BBC93 tracks and is also
in (perhaps better) agreement with the data. The edges of the MS and blue loop
evolutionary phases are less prominent than in the previous case thus allowing
a better reproduction of the smooth stellar distribution in the observational
C-M diagram. The model assumes again a moderately steep ($\alpha$=2.5) IMF
and two episodes of SF: one started 1 Gyr ago and lasted 8.5$\times10^8yr$ with
a SFR = $1.0 \times 10^{-2}M_{\odot}yr^{-1}$ and the second started
1.45$\times10^8yr$ ago and stopped 3$\times10^6yr$ ago with a lower SFR =
$5.1 \times 10^{-3}M_{\odot}yr^{-1}$.
\par
Finally, the model of Fig.12(c)
is based on the BBC86 tracks. Its stellar distribution is less satisfactory,
mostly because the blue edge of the blue loop phase is overpopulated and
because the red giants and supergiants do not reach large enough B-V.
The most consistent C-M diagram is obtained with $\alpha$=2.35
and two episodes of SF: the first started 1 Gyr ago and lasted
9$\times10^8yr$ with SFR = $1.9 \times 10^{-3}M_{\odot}yr^{-1}$ and the second
started 9.5$\times10^7yr$ ago and stopped 3$\times10^6yr$ ago (but, in this
case, allowing the SF to proceed till now does not significantly worsen the
agreement) with a larger SFR = $3.9 \times 10^{-3}M_{\odot}yr^{-1}$.
\par
The luminosity functions corresponding to the models of Fig.12 are shown in
Fig.13 together with the observed data. The top panel of Fig.13 represents
the global LF and the bottom panel the MS LF. As for Region A, for a more
direct comparison with the data, the synthetic global LFs include the
foreground stars. According to all our models,
objects in this region can be considered as safe MS stars only if their color
is -0.30$\leq$ B-V $\leq$0.20. All the three theoretical MS LFs are in
agreement
with the bright portion of the data. Fainter than V$\simeq$21.5 the BBC86
models appear in better agreement, but the uncertainty related to the higher
incompleteness at faint magnitudes weakens any choice between the stellar sets
of tracks. The global LFs are all in agreement with the empirical distribution.
\par
As already noticed for Region A, despite the different features of the three
sets of stellar models and the corresponding uncertainties, the simulations
provide consistent results. Independently of the adopted set of tracks, all
our best reproducing models indicate that in Region C of NGC 6822 the IMF has
been moderately steep and the SF preferably discontinuous.
\par
{}From the numerous simulations performed to test the effects of the various
assumptions on the SF and IMF, we have derived the following indications
on the history of Region C. A SFR constant over the last 1 Gyr is inconsistent
with the data, indipendently of the adopted IMF and set of tracks, because
it predicts an overpopulated MS and/or too many and too luminous bright stars.
Stopping the activity a few Myr ago in this field does not help to improve the
agreement (in some cases, it actually worsens it).
\par
A SFR exponentially decreasing with time does not provide satisfactory results
because it predicts too many bright post-MS stars. In addition, the appearance
of the MS strongly depends on the assumed IMF: with slopes flatter than
$\alpha\simeq$2.35 the MS is overpopulated in its brighter portion and with
steeper slopes it is generally underpopulated. The only exception to this
rule is the case of the BBC93 tracks and exponentially decreasing SFR
with average value $9 \times 10^{-3}M_{\odot}yr^{-1}$ which, combined with
an IMF slope $\alpha$=2.35-2.5, provides a stellar population consistent with
that observed in Region C (see the diagram in Fig.14 and the LFs in Fig.13).
\bigskip\noindent
6.~RESULTS
\medskip\noindent
{}From all the models discussed in the previous section, we have derived that
the two tested fields (Regions A and C) of NGC 6822 have had rather different
evolutions in the last 1 Gyr. Region A has had a moderately flat IMF
($2.0\leq\alpha\leq2.35$) and a fairly continuous SF (possibly decreasing
with time or interrupted for very short time intervals, depending on the
adopted set of stellar evolution models). Region C, instead, has had a
moderately steep IMF ($2.35\leq\alpha\leq2.6$) and a discontinuous SF.
Both results are consistent with those derived by Hodge (1980) from an
analysis of the stellar clusters and associations in this galaxy. If we
extrapolate the IMF over the range 0.1$\leq M/M_{\odot} \leq$120 and normalize
to the size of the observed region, the average SFR in Region A lies in the
range $(1 - 7) \times 10^{-8}M_{\odot}yr^{-1}pc^{-2}$ and the average rate
in Region C is in the higher range
$(6 - 15) \times 10^{-8}M_{\odot}yr^{-1}pc^{-2}$.
These rates are at least one order of magnitude higher than the SFR in the
solar neighbourhood ($\sim 7\times 10^{-9}M_{\odot}yr^{-1}pc^{-2}$) and
close to those attributed to the Large and Small Magellanic Clouds
(see e.g. Lequeux 1994).
\par
Thus, Region A has generated a stellar population with a lower total number of
stars than Region C because of the lower SFR, but a larger fraction of massive
bright objects because of the flatter IMF. This scenario explains the
different features of the observational C-M diagrams and luminosity functions
presented in Sections 4 and 5. In fact, the higher recent activity of star
formation  has generated in Region C not only a larger total number of stars,
but also an obvious larger fraction of MS objects. On the other hand, its
steeper IMF has made the MS more populated by less massive stars,
leading to the steeper MS LF derived from our observations. Had we studied
our regions with a simpler method, e.g. with a classic isochrone fitting,
we would not have found to which different combinations of SFR and IMF
the observed differences of the two stellar populations are due.
\par
The different color distribution of the examined fields in NGC 6822 seems to
be a simple color shift with no other strong differences implied (for instance
in the relative number or brightness of the red giants). We suggest that the
observed color shift
reflects the high differential reddening attributed to the galaxy by several
authors (cfr. Section 3) and not an intrinsic origin, like metallicity and/or
age of the three stellar populations.
\par
In the vast majority of our simulations in agreement with the data, we have
found that the star formation is currently inactive, specially in Region C
where no good model predicts an on going SF in the last 3 Myr. We must recall,
however, that our three fields were intentionally chosen for their lack of
HII regions and the large number of HII regions discovered by Hodge {\it et
al.}
(1988) in NGC 6822 indicates that this galaxy does currently form stars
elsewhere.
\par
As far as the stellar evolution models are concerned, we have seen in the
previous section that the two most recent sets provide results in better
overall agreement with the data. The older tracks (BBC86) predict in fact
a too large fraction of objects at the blue edge of the He burning loops.
Besides, they provide
a color extension of the stellar distribution narrower than the observed
one (but this might be due to its metallicity being slightly lower than that
attributed to NGC 6822). The models with moderate overshooting (BBC93) are
perhaps more
appropriate than those with classic treatment of convection (BBC93CL), but
this is more a feeling than a firm conclusion.
\bigskip\noindent
7.~GENERAL DISCUSSION
\medskip\noindent
Since NGC 6822 is the last galaxy of our proposed sample, we can draw here
some general considerations on the results of our project. All the observations
were made with the same telescope and instrumental setup, all the data reduced
with the same procedure and degree of accuracy, and all the simulations
performed with the same method and sets of stellar evolution models. We are
then confident that all the galaxies of the sample have been treated
homogeneously. This implies that the individual derived quantities (e.g. SFRs,
IMF slopes, etc.) may be uncertain, but the relative trends from one field to
another can be taken as real.
\par
All our galaxies are classified as dwarf irregulars and, indeed, they all have
small sizes (a few kpc across) and irregular shapes. They share the general
characteristics of the objects belonging to this morphological class of being
metal poor (Z$\simeq$0.001), with large gas contents and with young stellar
populations. With our method we have found that they share in addition a common
regime of {\it gasping} star formation but also show significant differences.
\par
{}From all the simulations performed on the various examined fields of our
sample
galaxies, we have inferred the evolutionary scenario summarized in Table 4.
Despite their small sizes, only two (Sextans B and DDO 210) of the systems
appear to be homogeneously populated by the same sort of stars. The other
three galaxies show different populations in different regions. The most
striking example of dishomogeneity is WLM where the most luminous stars
of the blue plume in Region 1 are two magnitudes brighter than those in Region
2
(Ferraro {\it et al.} 1989). This corresponds to a difference of 50-100 Myr in
the age of the younger objects of the two regions. It is apparent then that one
cannot treat these galaxies as homogeneous bodies with a single stellar
population. At the same time, in order to derive their global properties,
one needs to sample the stellar population of the whole galaxy.
\par
The IMF is also varying from one galactic region to another, but we do not
find evidence for large differences.
Our method is not sensitive to variations of the lower mass
cutoff of the IMF, since low mass stars are not detected in our frames, and
can discriminate only between upper mass cutoffs different by 2-3 tens of
solar masses. In fact, due to the small number of luminous stars observed in
each field and to the corresponding statistical fluctuations, simulations
assuming an upper mass limit of 100 M$_{\odot}$ predict number and
luminosity of the brightest stars not significantly different from those
assuming an upper mass limit of, say, 80 M$_{\odot}$. Vice versa, our method
can safely evidentiate variations in the IMF slope, because this affects
the relative numbers of stars at all the observed magnitudes and colors and
the corresponding statistics is more reliable. We have simulated each observed
field assuming different slopes, varying in the range 1.2$\leq\alpha\leq$3.3
(where 1.2 is the value at Z=0.001 corresponding to the metallicity dependent
slope proposed by Melnick 1987, and 3.3 is the slope in the massive star range
proposed by Tinsley 1980 for the solar neighbourhood). In Table 4 we list
the average slopes providing results in better agreement with the
observed data in each region.
The most popular slope turns out to be
Salpeter's $\alpha$=2.35, since only a couple of regions can be reproduced
also with a flatter slope and only a couple also with a steeper slope.
Salpeter's IMF is currently considered too flat for the solar neighbourhood
and for spiral galaxies in general, therefore our result may indicate that
low metallicity galaxies, like dwarf irregulars, have flatter IMFs as suggested
by some authors (e.g Melnick 1987 and references therein). However, the data
of only one of our observed fields are consistent with $\alpha$ as flat as
1.2, thus showing that in general the metallicity dependence of the IMF,
if any, must be rather shallow.
\par
As far as the star formation is concerned, the most important result of our
project is that all the studied fields appear to have had a sort of
discontinuous regime that we have named {\it gasping}. Not the strong
discontinuity attributed (e.g. Matteucci \& Tosi 1985, Pilyugin 1993,
Marconi, Matteucci \& Tosi 1994) to Blue Compact galaxies where a very few
short bursts of intense SF, separated by quiescent phases some Gyr long,
have taken place throughout the galaxy lifetime, but long episodes of,
sometimes modest, activity separated by short intervals.
This {\it gasping} regime may appear more or less discontinuous depending
on the examined field and on the adopted method of analysis, and is fairly
reminiscent of the irregular and discontinuous SF activity proposed by
Hodge (1980) for NGC 6822 and IC 1613. From this point of view, dwarf
irregulars
show an intermediate behaviour, between the extreme discontinuities of Blue
Compact galaxies and the steady state of giant irregulars.
\par
It is worth noticing that in all the studied regions we have found that the
best reproducing models predict no SF activity in the last few Myr. In the
cases of NGC 6822 and NGC 3109 this may result from the selection effect of
not observing fields containing HII regions, but in the cases of Sextans B
and DDO 210 the entire galaxies have been observed and we can then conclude
that no SF processes are currently taking place there.
\par
Finally, in most cases the SFR derived for the recent episode is quite
different (usually lower) from that derived for the previous one. This,
in our opinion, supports the hypothesis of discontinuous SF regime. In
addition,
the average SFRs derived for the sample galaxies may be fairly different from
each other. DDO 210 has the lower average rate in the last 1 Gyr with its
$(0.2 - 1.6) \times 10^{-8}M_{\odot}yr^{-1}pc^{-2}$ and a factor of ten lower
rate in the last 50 Myr. NGC 6822 shows the higher rate with its
$(0.5 - 1) \times 10^{-7}M_{\odot}yr^{-1}pc^{-2}$, and
Sextans B and NGC 3109 lie in between with an average recent rate of
$(0.3 - 3.5) \times 10^{-8}M_{\odot}yr^{-1}pc^{-2}$ and
$(0.3 - 8.3) \times 10^{-8}M_{\odot}yr^{-1}pc^{-2}$ respectively.
These values are quite consistent with those quoted in the literature for
other irregulars. For instance, Aparicio et al. (1987) find that in Sextans
A the average rate, transformed into our units, is about
$1.7 \times 10^{-8}M_{\odot}yr^{-1}pc^{-2}$, and Lequeux (1994) quotes
a SFR of several times $10^{-8}M_{\odot}yr^{-1}pc^{-2}$ for both Magellanic
Clouds. The large differences in the average star formation rates derived for
our examined galaxies suggest that, even if
they share a common SF regime, these galaxies might be in different physical
conditions. On the other hand, and despite the obvious statistical
uncertainties on such a small sample of objects, we notice as a general trend
that the galaxies
with lower inferred SFR are the smaller ones and those with larger SFR are the
bigger ones (cfr Greggio 1994). This is in agreement with the Stochastic Self
Propagating Star Formation theory by Gerola, Seiden \& Schulman (1980) that
predicts higher probabilities of triggering the star forming processes in
systems with larger number of cells, that is with larger size.
NGC3109 is however an exception to this rule, as its optical size is large
(intermediate between the SMC and LMC), and yet its SFR is significantly lower
than that inferred for NGC 6822, a much smaller galaxy. Although our observed
fields cover only a fraction of these two galaxies, given their size and
location over the main body of the systems, we believe that the derived C-M
Diagrams (and the corresponding SFRs) fairly represent their average stellar
populations. From de Vaucouleurs
{\it et al.} (1993), with our adopted distance moduli the diameters of the two
galaxies turn out to be $\sim$7 and $\sim$3 Kpc. Therefore, either there are
strong fluctuations around the average trend predicted by the SSPSF theory,
or the recent SFR does not trace well the overall activity.
Anyway, our sample contains too few galaxies and we need to analyze more
objects in order to test the SSPSF theory, possibly sampling the SF activity
over longer lookback times.
\par
In Papers I and II we have shown that a simple extrapolation to the whole
galaxy lifetime of the SF regime derived with our method for the last 1 Gyr
leads to overestimating the metallicity and to underestimating the gas
content in these systems. Marconi, Matteucci and Tosi (1994, in preparation)
are computing more detailed chemical evolution models for the galaxies of
our sample, assuming our results for the last 1 Gyr and various alternatives
for the previous epochs, to derive information on their previous histories.
To select the most plausible scenarios, the model predictions will be compared
with the observed chemical abundances of the different elements and with the
observed values of gas and total mass in each galaxy. The comparison of
the resulting pictures with those derived by other authors with different
approaches (e.g. Hodge 1994) will hopefully provide a clue to better understand
the evolution of irregular galaxies.

\bigskip
We are grateful to Harold Corwin for pointing out the error in the coordinates
of DDO 210 and to Christine Wilson for sending us her data files.
\vfill\eject
\parindent=-1truecm
\bigskip
REFERENCES
\medskip
\baselineskip=12pt
Aaronson, M., Cook, K.H., Norris, J. 1985, in Spectral Evolution of
Galaxies, C.Chiosi and A.Renzini eds (Reidel, Dordrecht Holland) p.171
\par
Aaronson, M., Mould, J., Cook, K.H., 1985, Astrophys.J. 291, L41
\par
Alcaino, G., Liller, W., Alvarado, F. 1987, A.J. 93, 1464
\par
Alongi, M., Bertelli, G., Bressan, A., Chiosi, C., Fagotto, F., Greggio, L.,
Nasi, E. 1993, Astron.Astrophys.Suppl.Ser. 97, 851
\par
Aparicio, A., Garc\'ia-Pelayo, J.M., Moles, M., Melnick, J. 1987,
Astron.Astrophys.
Suppl.Ser. 71, 297
\par
Azzopardi, M., Lequeux, J., Maeder, A. 1988, Astron.Astrophys. 189, 34
\par
Bertelli G., Bressan A., Chiosi C., Angerer K. 1986, Astron. Astrophys.
Suppl. Ser. 66, 191.
\par
Bonifazi, A., Fusi Pecci, F., Romeo, G., Tosi, M. 1990, M.N.R.A.S.,
245, 15
\par
Bresolin, F., Capaccioli, M., Piotto, G. 1993, A.J. 105, 1779
\par
de Vaucouleurs, G., de Vaucouleurs, A., Corwin, H.G.Jr, Buta, R.J.,
Paturel, G., \& Foqu\'e, P., 1991,
III Reference Catalogue of Bright Galaxies (Springer-Verlag, New York)
\par
Dennefeld, M. $\&$ Tammann, G.A. 1980, Astron.Astrophys. 83, 275
\par
Elias, J.H. and Frogel, J.A. 1985, Astrophys.J. 289, 141
\par
Fabbiano, G., Kim, D.W., Trinchieri, G. 1991, Ap.J.Suppl.Ser. 80, 531
\par
Ferraro, F.R., Fusi Pecci, F., Tosi, M., Buonanno, R. 1989, M.N.R.A.S.
241, 433
\par
Fisher, J.R. and Tully, R.B. 1975, Astron.Astrophys. 44, 151
\par
Freedman, W.L. 1985, Astrophys.J. 299, 74
\par
Freedman, W.L. 1988, A.J. 96, 1248
\par
Gallart, C., Aparicio, A., Chiosi, C., Bertelli, G., Vilchez, J.M. 1994,
Astrophys.J.Let. 425, L9
\par
Gerola, H., Seiden, P.E., Schulman, L.S. 1980, Astrophys.J. 242, 517
\par
Greggio L., 1984 in Observational Tests of the Stellar Evolution Theory,
IAU Symp.105, A. Maeder and A. Renzini eds (Dordrecht:Reidel), p. 329.
\par
Greggio L., 1994 in The Local Group: Comparative and Global Properties,
3rd CTIO/ESO Workshop, in press
\par
Greggio, L., Marconi, G., Tosi, M., Focardi, P., 1993, A.J. 105,
894, Paper II
\par
Hoessel, J.G. 1986, in Luminous Stars and Associations in Galaxies, IAU
Symp.116, C.W.H. de Loore, A.J.Willis \& P.Laskarides eds (Reidel Dordrecht
Holland), p.439
\par
Hoessel, J.G. \& Anderson, N. 1986, Astrophys.J.Suppl.Ser. 60, 507,
HA
\par
Hodge, P.W. 1980, Astrophys.J. 241, 125
\par
Hodge, P.W. 1994, in Dwarf Galaxies, G.Meylan \& P.Prugniel eds (ESO,
 Garching FRG) in press
\par
Hodge, P.W., Kennicutt, R.C.Jr., Lee M.G. 1988, PASP 100, 917
\par
Hubble, E.R. 1925, Astrophys.J. 62, 409
\par
Humphreys, R. 1980, Astrophys.J. 238, 65
\par
Hunter, D.A. $\&$ Gallagher, J.S.III 1986, PASP 98, 5
\par
Kayser, S.E. 1967, A.J. 72, 134
\par
Kennicutt, R.C.Jr. 1992, in Star Formation in Stellar Systems, G.
Tenorio-Tagle,
M. Prieto $\&$ F. S\'anchez eds, (Cambridge University Press, UK), p. 191
\par
Klein, U. \& Grave, R. 1986, Astron.Astrophys. 161, 155
\par
Landolt, A.U. 1983, A.J. 88, 439
\par
Lequeux, J. 1994, in Dwarf Galaxies, G.Meylan \& P.Prugniel eds (ESO,
 Garching FRG) in press
\par
Marconi, G., Focardi, P., Greggio, L., Tosi, M. 1990, Astrophys.J.
360, L39
\par
Marconi, G., Matteucci, F., Tosi, M. 1994, MNRAS in press
\par
Mateo, M. 1988, Astrophys.J. 331, 261
\par
Matteucci F. \& Tosi, M. 1985, MNRAS 217, 391
\par
McAlary, C.W., Madore, B.F., McGonegal, R., McLaren, R.A., Welch, D.L. 1983,
Astrophys.J. 273, 539, McA
\par
Melnick, J. 1987, in Stellar Evolution and Dynamics in the Outer Halo
of the Galaxy, M.Azzopardi \& F.Matteucci eds (ESO Garching FRG), p.589
\par
Pagel, B.E.J., Edmunds, M.G., Smith, G. 1980, MNRAS 193, 219
\par
Pilyugin, L.S. 1993, Astron.Astrophys. 272, 42
\par
 Romeo G, Bonifazi A, Fusi Pecci F., Tosi M., 1989, MNRAS, 240,459
\par
Rufener, F. 1986, Astron.Astrophys. 165, 275
\par
Stetson, P.B. 1987, PASP 99, 191
\par
Stetson, P.B. and Harris, W.E. 1988, A.J. 96, 909
\par
Tinsley, B.M. 1980, Fund.Cosmic Phys. 5, 287
\par
Thuan, T.X. \& Gunn, J. 1976, PASP 88, 543
\par
Tosi, M. 1994, in Dwarf Galaxies, G.Meylan \& P.Prugniel eds (ESO,
 Garching FRG) in press
\par
Tosi, M., Greggio, L., Marconi, G., Focardi, P. 1991, A.J.
102, 951, Paper I
\par
van den Bergh, S. \& Humphreys, R. 1979, A.J. 84, 604\par
\par
Wilson, C. 1992, A.J. 104, 1374
\par\noindent
\vfill\eject
Table 1: Local Group irregulars in our program
\eject\par
\parindent=1truecm
\medskip\noindent
\centerline {{\bf Table 1.} Local Group irregulars in our program.}
{\eightrm
\baselineskip=10pt
\tabskip=1em plus.7em minus.7em
\def\mrule{\noalign{\vskip6pt\hrule\vskip6pt}}
\mrule
\noindent
\halign to
\hsize{#\hfil&#\hfil&#&#\hfil&#&\hfil#&#&
\hfil#&#&\hfil#\hfil&#&\hfil#\hfil&#&\hfil#\hfil&#&\hfil#\hfil\cr
{}~~~~~~~~~~Name&~~~R.A.&&~~~~~DEC&&l~~~&&b~~~&&(m$-$M)$_o$&&NF&&detected&&selected\cr
\mrule
DDO ~~70 (Sextans B)&09 57 23&&+05 34 07&&233.2&&+43.8&&25.6 &&2&&2434&&1300\cr
DDO 209 (NGC 6822) &19 42 07&&$-$14 55 01&&~25.3&&$-$18.4&&23.5
&&3&&2031&&1772\cr
DDO 210            &20 44 08&&$-$13 02 00&&~34.1&&$-$31.4&&28?&&2&&1247&&
633\cr
DDO 221 (WLM)      &23 59 23&&$-$15 44 06&&~75.9&&$-$73.6&&25.0 &&3&&2513&&
$-$\cr
DDO 236 (NGC 3109) &10 00 48&&$-$25 55 00&&262.1&&+23.1&&25.7
&&3&&6430&&2605\cr
}
\mrule }
\medskip\par
\eject\par
Table 2: Journal of observations
\eject\par
Table 3a: Principal parameters of the selected stars of NGC 6822: Region A
\medskip\par
Table 3b: Principal parameters of the selected stars of NGC 6822: Region B
\medskip\par
Table 3c: Principal parameters of the selected stars of NGC 6822: Region C
\eject\par
Table 4: Evolutionary scenario resulting from our project
\eject\par
\bigskip\centerline {{\bf Table 4.} Evolutionary scenario resulting from our
project.}
\par
{\eightrm
\baselineskip=7pt
\tabskip=1em plus.8em minus.8em
\def\mrule{\noalign{\vskip6pt\hrule\vskip6pt}}
\mrule
\noindent
\halign to
\hsize{#\hfil&\hfil#\hfil&#&#&\hfil#\hfil&#&\hfil#\hfil&\hfil#\hfil\cr
Name&Homogeneity&&&IMF&&SF type&SFR\cr
\mrule
Sextans B &yes &&     & 2.6 && gasps stopped & SFR$_{young}\simeq$0.5
SFR$_{old}$ \cr
\mrule
          &    &&Reg.A& 2.2 && gasps stopped  & SFR$_{young}\simeq$0.5
SFR$_{old}$ \cr
          &    &&     & 2.35&& const stopped  & SFR$_{young}\simeq$1.0
SFR$_{old}$ \cr
NGC 6822  &no  &&$---$&$---$&&$-------$&$-----------$ \cr
          &    &&Reg.C& 2.6 && gasps stopped   & SFR$_{young}\simeq$0.4
SFR$_{old}$ \cr
          &    &&     & 2.35&& const stopped? & SFR$_{young}\simeq$0.2
SFR$_{old}$ \cr
\mrule
DDO 210   &yes &&     & 2.35&& gasps stopped   & SFR$_{young}\simeq$0.1
SFR$_{old}$ \cr
\mrule
          &    &&Reg.1& 2.35&& gasps stopped   & SFR$_{young}\simeq$2.0
SFR$_{old}$ \cr
WLM       &no  &&$---$&$---$&&$-------$&$-----------$ \cr
          &    &&Reg.2& 2.35?&& stopped 5-10 Myr ago  &\cr
\mrule
          &    &&Reg.A& 1.2 && gasps stopped   & SFR$_{young}\simeq$0.3
SFR$_{old}$ \cr
          &    &&     & 2.35&& gasps stopped   & SFR$_{young}\simeq$2.0
SFR$_{old}$ \cr
NGC 3109  &no  &&$---$&$---$&&$-------$&$-----------$ \cr
          &    &&Reg.C& 2.35&& gasps stopped   & SFR$_{young}\simeq$3.0
SFR$_{old}$ \cr
}
\mrule
}

\vfill\eject\noindent
{\bf FIGURE CAPTIONS}
\medskip\noindent
Figure 1: Enlargement of the blue print of the ESO Sky Survey with
our observed fields on NGC~6822.
\medskip\noindent
Figure 2: Plots of the residuals for the standard stars used for
the calibration.
\medskip\noindent
Figure 3: Distribution with magnitude of the photometric errors derived
with DAOPHOT for the B, V and R deepest frames of Region A.
\medskip\noindent
Figure 4: Completeness fraction in our deepest B (dotted line), V (solid
line) and R (dashed line) frames of Regions A, B and C.
\medskip\noindent
Figure 5: Differences between our magnitudes and those derived by HA (top
panels) and by Wilson (1992, bottom panels) for the common stars.
\medskip\noindent
Figure 6: Color-Magnitude diagram (V, B-V) for the stars with photometric
error smaller than 0.1 mag in our three fields of NGC 6822.
\medskip\noindent
Figure 7: Color-Magnitude diagram (V, B-V) of the objects in the external
field of NGC~6822.
\medskip\noindent
Figure 8: Differential luminosity function for all the objects selected in
the three observed regions (top panel) and for only MS stars (bottom
panel, see text for the MS selection criterion). The dotted lines refer to
Region A, the dashed lines to Region B and the solid lines to Region C.
\medskip\noindent
Figure 9: Synthetic C-M diagrams for Region A based on the best models
with the BBC93CL (top), the BBC93 (center) and the BBC86 (bottom) stellar
tracks.
\medskip\noindent
Figure 10: Differential luminosity function for all (top) and only MS (bottom)
stars in Region A. The dots represent the observed values corresponding to
stars
selected in the B and V bands, for homogeneity with Region C. The dashed lines
correspond to the model of panel (a) of Fig.9, the thick solid lines to that of
panel (b) and the dotted lines to that of panel (c). The thin solid line
corresponds to the model of Fig.11.
\medskip\noindent
Figure 11: Synthetic C-M diagrams for Region A based on the BBC93CL tracks
and with exponentially decreasing SFR.
\medskip\noindent
Figure 12: Synthetic C-M diagrams for Region C based on the best models
with the BBC93CL (top), the BBC93 (center) and the BBC86 (bottom) stellar
tracks.
\medskip\noindent
Figure 13: Differential luminosity function for all (top) and only MS (bottom)
stars in Region C. The dots represent the observed values. The dashed lines
correspond to the model of panel (a) of Fig.12, the thick solid lines to that
of
panel (b) and the dotted lines to that of panel (c). The thin solid line
corresponds to the model of Fig.14.
\medskip\noindent
Figure 14: Synthetic C-M diagrams for Region C based on the BBC93 tracks
and with exponentially decreasing SFR.
\bye